\documentclass{aa} 
\usepackage{graphicx}
\usepackage{txfonts}
\usepackage{color}
\usepackage{float}
\newcommand{\cmthree}{cm$^{-3}$}

\newcommand{\um}{$\mu$m}                                 
\newcommand{\acena}{$\alpha \, {\rm Cen\,A}$}          
\newcommand{\acenb}{$\alpha \, {\rm Cen\,B}$}                          %
\newcommand{\acen}{$\alpha \, {\rm Cen}$}                              %
\newcommand{\about}{$\sim$}                       
\newcommand{\powten}[1]{10$^{#1}$}
\newcommand{\amin}{$^{\prime}$}                   
\newcommand{\asec}{$^{\prime \prime}$}
\newcommand{\adeg}{$^{\circ}$}

\newcommand{\radot}[4]{\mbox{#1$^{\rm h}$#2$^{\rm m}$#3$\stackrel {\rm s}{_{\bf\cdot}}$#4}}  

\newcommand{\decdot}[4]{\mbox{#1$^{\circ}$ #2$^{\prime}$ #3$\stackrel {\prime \prime}{_{\bf \cdot}}$#4}}

\newcommand{\asecdot}[2]{\mbox{#1$\stackrel {\prime \prime}{_{\bf \cdot}}$#2}}

\begin{document}

\title{ALMA's view of the Sun's nearest neighbours}
\subtitle{The submm/mm SEDs of the  $\alpha$ Centauri binary and a new source}

   \author{
                		R. Liseau\inst{1}
          \and
          		V. De la Luz\inst{2}             		
	\and
			E. O'Gorman\inst{3}	  
         \and
                		E. Bertone\inst{4}
         \and
                        	M. Chavez \inst{4}
	\and
			F. Tapia\inst{2}	                      
          }

   \institute{Department of Earth and Space Sciences, Chalmers University of Technology, Onsala Space Observatory, SE-439 92 Onsala, Sweden,
                                 \email{rene.liseau@chalmers.se}    		 																
     \and       
                SCiESMEX, Instituto de Geofisica, Unidad Michoacan, Universidad Nacional Autonoma de Mexico, Morelia, Michoacan, Mexico 		  
      \and 
          	Dublin Institute for Advanced Studies, Astronomy and Astrophysics Secton, 31 Fitzwilliam Place Dublin 2, Ireland 					  
    \and                               
                Instituto Nacional de Astrof\'{\i}sica, \'{O}ptica y Electr\'{o}nica (INAOE), Luis Enrique Erro 1, Sta. Mar\'{i}a Tonantzintla, Puebla, Mexico       
           }
                
   \date{Received ; accepted}
  \abstract
   {The precise mechanisms that provide the non-radiative energy for heating the chromosphere and the corona of the Sun and other stars are at the focus of intense contemporary research.}
   {Observations at submm/mm wavelengths are particularly useful to obtain information about the run of the temperature in the upper atmosphere of Sun-like stars. We used the Atacama Large Millimeter/submillimeter Array (ALMA) to study the chromospheric emission of the $\alpha$\,Centauri binary system in all six available frequency bands during Cycle 2 in 2014-2015.}
   {Since ALMA is an interferometer, the multi-telescope array is particularly suited for the observation of point sources. With its large collecting area, the sensitivity is high enough to allow the observation of nearby main-sequence stars at submm/mm wavelengths for the first time. The comparison of the observed spectral energy distributions with theoretical model computations provides the chromospheric structure in terms of temperature and density above the stellar photosphere and the quantitative understanding of the primary emission processes.}
   {Both stars were detected and resolved at all ALMA frequencies. For both \acena\ and B, the existence and location of the temperature minima, firstly detected from space with {\it Herschel}, are well reproduced by the theoretical models of this paper.  For \acenb, the temperature minimum is deeper than for A and occurs at a lower height in the atmosphere, but for both stars, $T_{\rm min}/T_{\rm eff}$ is consistently lower than what is derived from optical and UV data. In addition, and as a completely different matter, a third point source was detected in Band 8 (405\,GHz, 740\,\um) in 2015. With only one epoch and only one detection, we are left with little information regarding that object's nature, but conjecture that it might be a distant solar system object.}
   {The submm/mm emission of the \acen\ stars is indeed very well reproduced by modified chromospheric models of the Quiet Sun. This most likely means that the non-radiative heating mechanisms of the upper atmosphere that are at work in the Sun are operating also in other solar-type stars.}
   \maketitle
%

\section{Introduction}

Outside the solar system, Alpha Centauri (\acen) is our nearest neighbour, only a little more than a parsec away ($\pi=\,$\asecdot{0}{742}). It is a double star, and its primary \acena\ has the same spectral type and luminosity class as the Sun, viz. G2\,V. The secondary, \acenb, is a somewhat cooler star, of spectral type K1\,V. Using asteroseismology, the age of the main-sequence stars  \acena\ and B has been determined to $4.85\pm 0.5$\,Gyr  by \citet{thevenin2002}, whereas statistical methods resulted in estimates of 8 to 10\,Gyr, depending on the method used, the Ca\,II\,$R^{\prime}_{\rm HK}$ index or the X-ray luminosity, respectively \citep[see, e.g.,][and references therein]{eiroa2013}. 

The proximity of \acen, the similarity of A, and the differences of B, compared to the Sun provide an excellent opportunity to study the stellar-solar relationship, as the understanding of the physics of the Sun and the stars is an iterative process that provides feed-back in both directions. For instance, an outstanding problem of modern solar physics is the heating of the outer atmospheric layers, i.e., of the chromosphere and the corona \citep{wedemeyer2007}. A few hundred kilometers above the solar photosphere, the temperature gradient changes sign at the location of the temperature minimum. From early theoretical models of the chromosphere, this phenomenon was already found also for \acena\ and B \citep[and in addition, for $\alpha$\,Boo and $\alpha$\,CMi:][]{ayres1976}. The primary observables were the wings of optical and UV resonance lines, e.g. Ca\,II\,H\&K and Mg\,II\,h\&k, the cores of which are formed higher up in the chromosphere. In addition, high temperature tracers also include high ionization lines and continua in the UV from the transition region and radio emission and X-rays from the corona. 
\begin{table*}
\caption{Positions of \acena\ and B with ALMA in  Right Ascension and Declination (ICRS J\,2000.0) }             
\label{pos}      
\centering          
\begin{tabular}{ c cccccc}     
\hline\hline \\ 
\smallskip
\smallskip   
		& Date		& Start UTC		& End UTC	& \acena				                         & \acenb			           			& Synthesized  Beam\\
		& yyyy-mm-dd	& hh min sec 		& hh min sec 	& hh\,mm\,ss.s\hspace*{1cm}\adeg\,\,\amin\,\,\asec	& hh\,mm\,ss.s\hspace*{1cm}\adeg\,\,\amin\,\,\asec 	&  $a$\asec\ $\times\,\,b$\asec\hspace*{0.5cm} $PA$\adeg \\
\hline \\
B3		& 2014-07-03	& 00 47 20.4 		& 01 38 19.4 	&   14\, 39\, 28.893\,\,  $-60\, 49\, 57.86$	& 14\, 39\, 28.333\,\, $ -60\, 49\, 56.94$	& $1.81\times\,1.22$ \hspace*{0.3cm}19	  \\
B7		& 2014-07-07	& 02 26 26.4   		& 02 44 53.8  	&   14\, 39\, 28.883\,\,  $-60\, 49\, 57.84$	& 14\, 39\, 28.325\,\, $ -60\, 49\, 56.91$	& $0.43\times\,0.28$ \hspace*{0.3cm}47	  \\
B9		& 2014-07-18	& 00 56 05.7   		& 01 26 49.4	&   14\, 39\, 28.870\,\,  $-60\, 49\, 57.83$	& 14\, 39\, 28.309\,\,  $-60\, 49\, 56.89$	& $0.22\times\,0.16$ \hspace*{0.3cm}36	   \\
B6		& 2014-12-16  	& 11 04 36.6       	& 11 18 34.2 	&   14\, 39\, 28.650\,\,  $-60\, 49\, 57.60$	& 14\, 39\, 28.120\,\, $-60\, 49\, 56.32$	& $1.64\times\,1.07$ \hspace*{0.3cm}71	   \\
B4		& 2015-01-18	& 13 35 24.5		& 13 59 40.8	&   14\, 39\, 28.624\,\,  $-60\, 49\, 57.63$	& 14\, 39\, 28.110\,\, $-60\, 49\, 56.27$	& $3.16\times\,1.67$ \hspace*{0.3cm}82	   \\
B8		& 2015-05-02	& 03 04 14.2		& 03 25 01.7	&   14\, 39\, 28.439\,\,  $-60\, 49\, 57.44$	& 14\, 39\, 27.934\,\, $-60\, 49\, 55.85$	& $0.77\times\,0.68$ \hspace*{0.2cm}$-70$  \\
\hline                
\end{tabular}
\end{table*}

\begin{table*}
\caption{ALMA flux density data for the \acen tauri binary}             
\label{data}      
\centering          
\begin{tabular}{c  l l l l l l }     
\hline\hline \\ 
\smallskip
\smallskip     
                         & \multicolumn{6}{c}{Primary beam corrected flux density, $S_{\!\nu} \pm \Delta S_{\!\nu}$ (mJy),  and signal-to-noise [S/N]}  \\
                  \cline{2-7}  \\                                                                                   		                    
	        & Band 9   		 		& Band 8               			& Band 7  						& Band 6    					&  Band 4    			& Band 3           \\         
		& 679\,GHz		    		& 405\,GHz			     	& 343.5\,GHz					& 233\,GHz 					& 145\,GHz		 	& 97.5\,GHz	 \\
		& 442\,\um 				& 740\,\um				& 873\,\um 					& 1287\,\um 					& 2068\,\um			& 3075\,\um	\\
\hline \\
A  & $107.2\pm 1.50$ [71]                     &  $35.32\pm 0.21$1 [168]             & $26.06\pm 0.19$ [137]                	&  $13.58\pm 0.08$ [170]                 	&   $6.33\pm 0.08$ [83]      & $3.37\pm 0.012$ [281]     \\
B  & \phantom{1}$57.6\pm 4.5$\,\,\,\,[13]   & $16.53\pm 0.19$\,\,\,\,\,[87]   & $11.60\pm 0.34$ [34]                   	& \phantom{1}$6.19\pm 0.05$ [124] 	&   $2.58\pm 0.08$ [34]      & $1.59\pm 0.02$\,\,\,\,\,[80]  \\    
\hline                
\end{tabular}
\end{table*}

The temperature minimum of \acen\ was directly observed in the far-infrared spectral energy distribution (SED) by \citet{liseau2013}. However, the far-infrared data did not resolve the binary in its individual components and the interpretation had to rely on photometry at shorter wavelengths. Observations with the Atacama Large Millimeter/submillimeter Array (ALMA) at three frequencies finally resolved the pair and the individual SEDs were spectrally mapped throughout the sub-millimeter (submm), up to 3\,mm \citep{liseau2015}. \acen\ was observed with three more ALMA bands during Cycle\,2. The stars themselves were unresolved and appeared as point sources to ALMA. With regard to the stellar-solar connection, these observations would refer to analogues of the Quiet Sun, for which the intensity is integrated over the solar disk.

The metallicity of \acen\ is slightly higher than that of the Sun, i.e. ${\rm [Fe/H]} = +0.24\pm 0.04$ \citep{torres2010}, a fact that could favor the existence of planets around the stars \citep[e.g.,][]{wang2015}. Examining a wealth of radial velocity data,  \citet{dumusque2012} announced the discovery of an Earth-mass planet around \acenb. That was however challenged by \citet{hatzes2013}, \citet{demory2015} and \citet{rajpaul2016} who were unable to confirm the existence of this object. 

Attempts to detect planets around \acen\ with direct imaging in the optical and the near infrared have hitherto been unsuccessful, see \citet{kervella2006,kervella2007} and Kervella et al. (2016, in preparation). At these wavelengths, any feeble planetary signal within several arcseconds from the stars would be totally  swamped by their overwhelming glare ($V$-magnitude = $-0.1$), alternatively be hidden behind the coronagraphic mask inside the inner working angle. This contrast problem would be naturally overcome for closeby faint objects with ALMA, an interferometer that for point sources in the reconstructed images generates a much cleaner point spread function (PSF), and our imaging results of \acen\ with ALMA are discussed toward the end of this paper.

The organization of this paper is as follows: Sect.\,2 reports the observations and the data reduction. Sect.\,3 briefly presents the results, which are discussed in Sect.\,4. We round off with our conclusions in Sect.\,5.

\begin{figure*}
  \resizebox{\hsize}{!}{
    \rotatebox{00}{\includegraphics{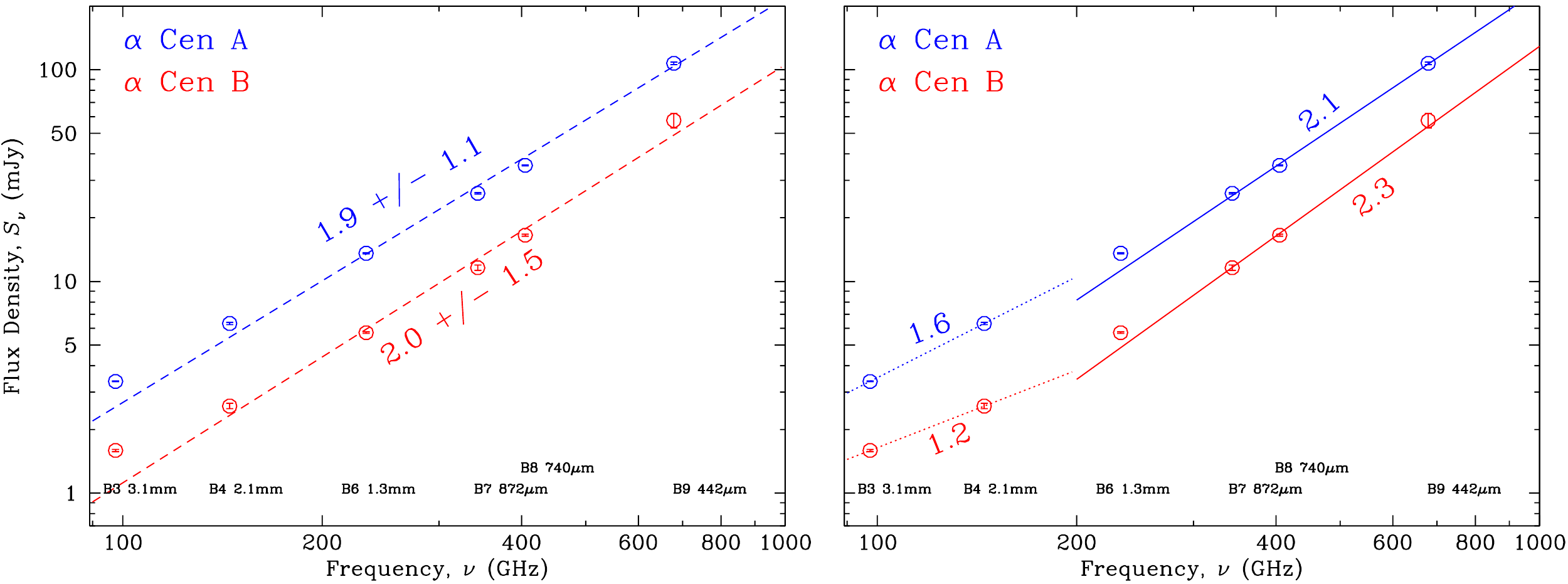}}
                        }
  \caption{Measurements of the flux density of \acena\ (blue circles) and of \acenb\ (red circles) with ALMA, with statistical $1 \sigma$ error bars inside the symbols.  {\bf Left:} Assuming that  $S_{\nu} \propto \nu^{\,\alpha}$, least-square fits to the Band\,3 to 9 flux densities are shown by dashed lines, with the power law exponent $\alpha = d\log S_{\!\nu}/d\log\nu$ shown next to them. {\bf Right:} Shown by the solid lines are fits, performed as in the left panel, to the data above, and by dotted lines below 200\,GHz (\about\,1.5\,mm). The ALMA bands, with their central wavelengths, are identified at the bottom of the figure.}
  \label{flux}
\end{figure*}

\begin{figure*}
  \resizebox{\hsize}{!}{
    \rotatebox{00}{\includegraphics{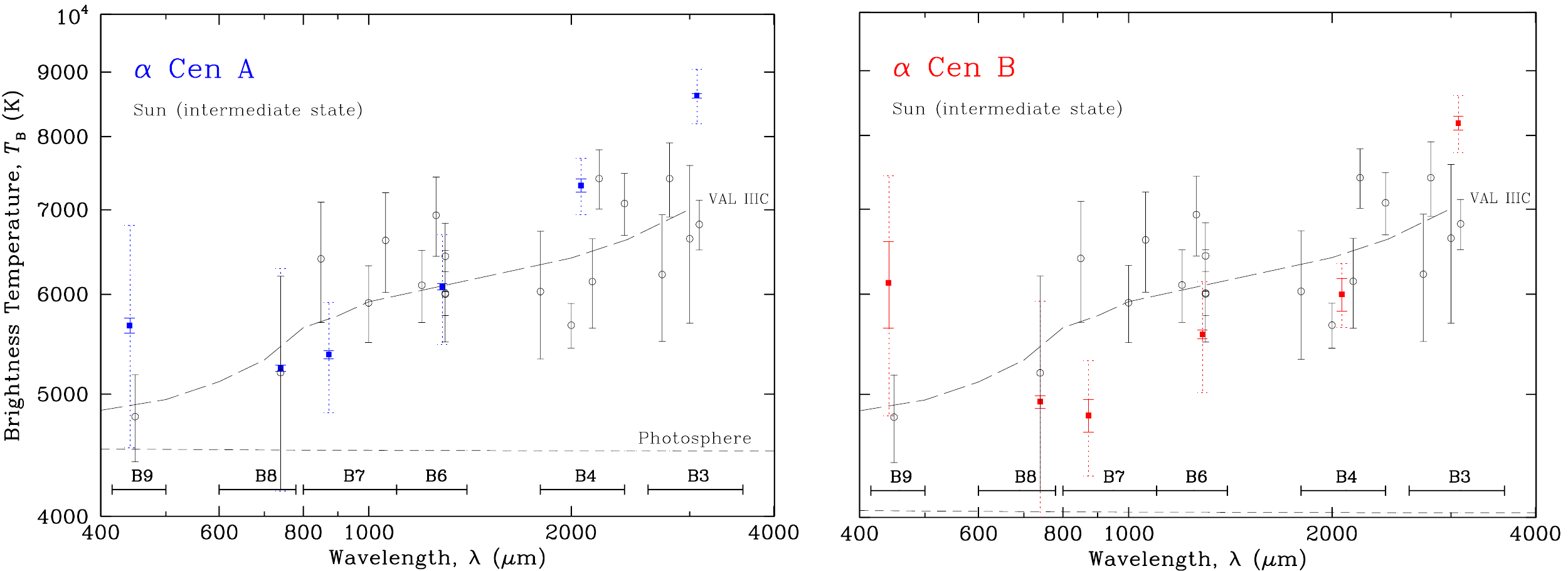}}
                        }
  \caption{Brightness temperature $T_{\rm B}$ in Kelvin at ALMA wavelengths $\lambda$ in \um, for Bands 3 to 9 of the G-star \acena\ (left, blue) and the K-star \acenb\ (right, red). In addition to the observational rms-errors (solid bars), the estimated absolute errors, including calibration uncertaities, are shown as dashed error bars. The stellar photospheres are represented by extrapolations to PHOENIX model atmospheres  of \citet{brott2005} for the stars' respective ($T{\rm eff}$, $\log g$, [Fe/H]) and are shown as black dashed lines. The ALMA bands are indicated below. A solar model chromosphere \citep[VAL IIIC,][]{vernazza1981} is shown as long dashes, with data for the Sun from \citet{loukitcheva2004} as black open circles.
   }
  \label{alma_ab}
\end{figure*}

\section{Observations and data reduction}

The binary \acena B was observed in all six ALMA continuum bands during the period July 2014 to May 2015 (Table\,\ref{data}). The field of view (primary beam) varied from about 10\asec\ for the shortest wavelength to about 1\amin\ for the longest. Similarly, the angular resolution (synthesized half power beam width) ranged from \asecdot{0}{2} to \asecdot{1}{5}. With angular diameters of \asecdot{0}{008} and \asecdot{0}{006} for A and B at 2\,\um\ \citep{kervella2003}, the stars were-point like to the ALMA interferometer in all wave-bands (cf. Table\,\ref{pos}). The ALMA program code is 2013.1.00170.S and the observations in Band\,3, 7 and 9 have already been described in detail by \citet{liseau2015} and will not be repeated here.

The observations in Band\,4, 6, and 8 were taken in the standard wideband continuum mode with 8\,GHz effective bandwidth spread over four spectral windows in each of the bands. The Band\,4 observations, taken on 2015 Jan 18 with 34 antennas, were centered on 145 GHz (2068\,\um), with $\sim$\,24 min of observing time with 5.5\,min on-source. The Band\,6 observations, taken on 2014 Dec 16 with 35 antennas, were centered on 233\,GHz (1287\,\um), with $\sim$\,14 min of observing time with $\sim$\,2 min on-source. Finally, the Band\,8 observations, taken on 2015 May 2 with 37 antennas, were centered on 405\,GHz (740\,\um), with $\sim$\,21 min of observing time with $\sim$\,7 min on-source.

The visibilities were flagged and calibrated following standard procedures using the CASA package\footnote{CASA is an acronym for {\it Common Astronomy Software Application.}} v4.2.2 for Band\,4 and 6, and v4.3.1 for Band\,8. The quasar J1617-5848 was used as complex gain calibrator in Band\,4 and 8, while J1408-5712 was used in Band\,6. The quasar J1427-4206 was used as bandpass calibrator in Band\,6 and 8, while J1617-5848 was used in Band\,4. Flux calibration was done using Ceres in Band\,4 when at 74\adeg\ elevation, while \acen\ was at 44\adeg. The quasar 1427-421 was used for flux calibration in Band\,6 when it was at 57\adeg\ elevation and \acen\ was at 46\adeg, while Titan was used in Band\,8 when it was at 50\adeg\ elevation and \acen\ was at 52\adeg.

Imaging was performed using natural weighting in Band\,4, 6, and 8 with one round of phase-only self-calibration was carried out on all three images to improve the rms noise. The synthesized beam sizes are listed in Table\,\ref{pos} and the primary beam-corrected flux densities and the rms noise per synthesized beam in the pointing center are listed in Table\,\ref{data}. We also imaged the ALMA spectral windows separately in each band to assess the spectral index within each band and the resultant flux densities for \acena\ and B are listed in Table\,A.1. and A.2., respectively.
\begin{table}

\caption{Stellar flux ratios and in-band (spw\,1 - spw\,4) spectral indices. $^*$ Too small bandwidth or too large errors. }             
\label{indices}      
\centering          
\begin{tabular}{c cc cc ccc }     
\hline\hline \\ 
\smallskip
\smallskip     
B	& $\lambda$		& $\nu$					& $S_{\!\nu}(B)/S_{\!\nu}(A)$	& $\alpha_{\alpha\,{\rm Cen\,A}}$ & $\alpha_{\alpha\,{\rm Cen\,B}}$ \\	
	&	(\um)			& (GHz)					&						& in-band			& in-band	\\
\hline    \\           
9$^*$	&  \phantom{1}442	& 679					& $0.54 \pm 0.044$		& $\cdots$		& $\cdots$		\\	
8		& \phantom{1}740	& 405					& $0.47 \pm 0.008$		& $1.3$			& $1.6$	\\
7$^*$	& \phantom{1}873	& \phantom{1}343.5			& $0.44 \pm 0.015$		& $\cdots$		& $\cdots$		\\
6		& 1287			& 233					& $0.46 \pm 0.007$		& $1.5$ 			& $0.9$	\\
4		& 2068			& 145					& $0.41 \pm 0.017$		& $1.8$ 			& $2.0$	\\
3		& 3075			&\phantom{1}\phantom{1}97.5  &$0.47 \pm 0.007$		& $1.7$			& $1.6$	\\
\hline  
\end{tabular}
\end{table}

\section{Results}

The binary system is well resolved at all frequencies. The J2000-coordinates for \acena\ and B on the observational dates are presented in Table\,\ref{pos}, together with the sizes of the synthesized beams (ellipses with semi-major axes $a$ and semi-minor axes $b$ in arcseconds) and their orientations (position angle $PA$ in degrees). The frequencies of the bands are given in Table\,\ref{data}, where the primary beam corrected flux densities, $S_{\!\nu}$, are reported together with their statistical errors. As can be seen, the signal-to-noise ratio, S/N, spans the range 10--100 for \acenb, and excels to nearly 300 for \acena. The absolute flux calibration is quoted in terms of goals\footnote{https://almascience.nrao.edu/documents-and-tools/cycle-2/alma-proposers-guide}, viz. better than 5\% for bands B\,3 and B\,4, better than 10\% for B\,6 and B\,7, and at best about 20\% for B\,8 and B\,9. These goals are shown for \acena\ and B in Fig.\,\ref{alma_ab}.

\subsection{Relative fluxes from 0.4 to 3.1\,mm}

The average flux ratio for the binary over the ALMA bands 3 through 9 is $[S_{\!\nu}({\rm B})/S_{\!\nu}({\rm A})]_{\rm ave}=0.464 \pm 0.051$ (Table\,\ref{indices}). This would be close to the ratio of their respective solid angles $(R_{\rm B}/R_{\rm A})^2=0.497\pm 0.003$, where the radii are those of  their interferometrically measured photospheric disks of uniform brightness \citep{kervella2003}. Comparison with the value for the range 0.09\,\um\ to 70\,\um, i.e., $0.44\pm 0.18$ \citep{liseau2013}, indicates an apparently remarkable constancy of the flux ratio over four orders of magnitude in wavelength, from the photospheric emission in the visible to that in the micro-wave regime.

\subsection{Spectral slopes of the SEDs}

A first order characterization of the emission mechanism(s) can be obtained from the spectral slope of the logarithmic SED. Assuming that  $S_{\nu} \propto \nu^{\,\alpha}$, linear regression \citep{press1986}\footnote{$\chi^2(a,\,b) = \sum_{i=1}^N [(y_i - a -bx_i)/\sigma_i]^2$, and $Q=\Gamma\,(\frac{N-2}{2},\,\frac{\chi^2}{2})$.} to the Band\,3 to 9 data results in a spectral index $\alpha_{\rm A,\,3-9}=1.92 \pm 1.06$ with a $\chi^2=0.015$ for \acena. For \acenb, the corresponding $\alpha_{\rm B,\,3-9}=1.97 \pm1.50$ and $\chi^2=0.033$,  see Fig.\,\ref{flux}. The goodness-of-fit is $Q=0.9999$ for both.  

This apparent constancy of the slope close to a value of two over the entire ALMA range, from 0.4 to 3.1\,mm, is perhaps surprising. A more careful inspection of the data reveals that the slopes at the shorter wavelengths appear marginally steeper, but that the long-wavelength data, not totally unexpected, seem to flatten out. Dividing the data into two sub-sets for both stars, i.e. below and above 1.5\,mm (200\,GHz), yields for the spectral indices of the \acena-SED $\alpha_{{\rm A},\,34}=1.6$ and  $\alpha_{{\rm A},\,69}=2.1$. Similarly, for \acenb, $\alpha_{{\rm B},\,34}=1.2$ and  $\alpha_{{\rm B},\,69}=2.3$ (Fig.\,\ref{flux}). In these cases, the formal fit errors are considerably larger for both \acena\ and B. However with regard to the fits in the left panel, the observed Band\,3 flux densities are in excess by more than $110\,\sigma$ for A and by more than $30\,\sigma$ for B. Therefore, the flattening of the SEDs towards lower frequencies is real. 

Observations at longer wavelengths would help to better constrain the run of the SED. Unfortunately, at declination south of $-60$\adeg\ the number of sensitive observing facilities is limited. \citet{trigilio2013} and (2014) proposed Australia Telescope Compact Array (ATCA) observations at 15\,mm (17\,GHz) and 16\,cm (2\,GHz). C.\,Trigilio privately communicated to us that both stars were recently detected at 17\,GHz. However, having no further information, we provide here our own flux estimates for ATCA observations of the binary $(S/N > 5)$. These are based on extrapolations beyond ALMA-Band 3 and the sensitivity specifications of the 6\,km compact array for the K-band (15\,mm) and C/X-band (4\,cm)\footnote{http://www.narrabri.atnf.csiro.au/observing/users\_guide/html/atug.html}, resulting in estimates of the $S/N = 54$ (0.27\,mJy) and 13  (0.13\,mJy) for \acena\ and $S/N = 26$ (0.04\,mJy) and 7 (0.02\,mJy) for \acenb, respectively. These values refer to 12\,hour on-source integrations (rms = 0.003\,mJy). The corresponding brightness temperatures are shown below, in Fig.\,\ref{stars_SED}.

Spectral indices for flux integrations over the individual bands are shown in Table\,\ref{indices}, except for Band\,9, where the fractional bandwidth is too small for meaningful measurement, and for Band\,7, where the relative errors are too large (negative slope within the band). Inside the individual bands, the data were collected through four spectral windows (spw; see Fig.\,\ref{flux_sw}), with the flux data for these provided in Appendix\,A. 

\begin{table*}
\caption{Brightness temperatures and chromospheric heights of \acena}             
\label{tb}      
\centering          
\begin{tabular}{ccc ccc cc}     
\hline\hline \\ 
\smallskip
\smallskip     
ALMA	& $\lambda$	&$\nu$					& $S_{\!\nu,\,\rm obs}$ 			 		&$S_{\!\nu,\,\rm phot}$			&$\Delta S_{\!\nu}$			& $h$		& $T_{\rm B}$ 		\\	
Band		&	(\um)		& (GHz)					& (mJy)								& (mJy)						& (mJy)					& (km)		& (K)				\\
\hline    \\           
3		& 3075		& \phantom{1}\phantom{1}97.5&  \phantom{1}\phantom{1}$3.37  \pm  0.01$	& \phantom{1}$1.76 \pm 0.09$		& \phantom{1}$1.61 \pm 0.09$	&2143:		& $8618 \pm 31$	\\
4		& 2068		& 145					&  \phantom{1}\phantom{1}$6.33 \pm   0.08$ 	& \phantom{1}$3.90 \pm 0.20$		& \phantom{1}$2.43 \pm 0.22$	&2140:		& $7316 \pm 88$	\\
6		& 1287		& 233					&  \phantom{1}$13.58 \pm  0.08$			&	$10.0 \pm 0.50$			& \phantom{1}$3.58 \pm 0.50$ 	&1180		& $6087 \pm  36$	 \\
7	& \phantom{1}873	& \phantom{1}343.5			&  \phantom{1}$26.06 \pm   0.19$			&	$21.9 \pm 1.10$			& \phantom{1}$4.16 \pm 1.11$	&\phantom{1}965& $5351 \pm  49$	\\
8	& \phantom{1}740	& 405					&  \phantom{1}$35.32 \pm  0.21$			&	$30.4 \pm 1.52$			& \phantom{1}$4.92 \pm 1.53$ &\phantom{1}950& $5242 \pm 55$	\\
9	& \phantom{1}442	& 679					& $107.20  \pm  1.50$ 					&	$85.3 \pm 4.26$			& 	$21.90 \pm 4.52$	 	&1050		& $5678 \pm  79$	\\	   
\hline  
\end{tabular}
\end{table*}


\section{Discussion}
\subsection{The stellar brightness temperatures}

The direct observation of the temperature minima of \acena B at far infrared wavelengths indicated a clear kinship with the Sun's chromosphere \citep{liseau2013,liseau2015}. At these wavelengths, the continuum opacity is dominated by inverse bremsstrahlung, with some contribution due to free-free H$^-$ processes \citep[e.g.,][]{dulk1985,wedemeyer2015}.

Fig.\,\ref{alma_ab} displays the observed spectral energy distributions of both stars in terms of their brightness temperatures\footnote{The brightness temperature, or radiation temperature, is the temperature of a blackbody that emits the same amount of radiation as the observed flux at a given frequency.} 
\begin{equation}
T_{\rm B}(\nu) = \frac { 2\,\pi\,\hbar\,\nu }{ k } \left [  \ln \left ( \frac{4\,\pi^2\,R^2_{\rm star}(1.0+h/R_{\rm star})^2\,\hbar\,\nu^3}{D^2\,c^2\,S\!_{\nu}} + 1\right ) \right ]^{-1}\,\,,
\end{equation}
where $R_{\rm star}$ is the stellar radius, $h$ the height at which the observed radiation originates, $\nu$ is the radiation frequency, $D$ is the distance to the source, $S_{\!\nu}$ is the observed flux density, and the other symbols have their usual meaning.

For the Sun, $h/R \sim 10^{-4}$, where $h$ refers to the height above the solar photosphere, where the optical depth in the visual $\tau_{5000}=1$ and $h=0$.  We assume similar $h/R$-values for the \acen\ stars and use their photospheric radii, i.e. $R_{\rm star}+h \sim R_{\rm star}$, where $R_{\rm star}$ refers to the values determined by \citet{kervella2003}. When $h \nu/k T \ll1$ (Rayleigh-Jeans regime), Eq.\,1 simplifies to
\begin{equation} 
T_{\rm B} \approx  \left ( \frac{D}{R_{\rm star}}\right )^2 \frac{c^2}{2\,\pi\,k\,\nu^2}\,S_{\!\nu}\,\,.
\end{equation}

Consequently, in the Rayleigh-Jeans regime (RJ), optically thick free-free emission (or {\it \emph{Bremsstrahlung}}) will behave as $S_{\!\nu} \propto \nu^2$, so that the spectral index, $\alpha = \Delta \log S_{\!\nu}/\Delta \log{\nu} = 2$ (Fig.\,\ref{flux}).  In that case, observed brightness temperatures correspond to actual physical temperatures. The data for the \acen\ stars reveal a positive temperature gradient, reminiscent of the solar chromosphere, and different frequencies probe the temperature stratification of the atmosphere. To determine the chromospheric height values $h$, requires a structure model of the atmosphere, that details the run of density and fractional ionization of the gas \citep[][and references therein]{delaluz2014,loukitcheva2015}.  

In Fig.\,\ref{a_sun}, $T_{\rm B}(\lambda)$ for the disk integrated \acena\ is compared with observed values for the Quiet Sun \citep{loukitcheva2004}.

\subsection{Theoretical model chromospheres for \acen}

The region close to the temperature minimum is optically thick in the FIR/submm \citep{liseau2015} which, as a consequence of the negative temperature gradient, limits our view to higher, cooler layers above the optical photosphere. Therefore, the received flux at a given frequency measures directly the temperature of the plasma at a particular atmospheric height. That can be used to construct analytically the temperature profile to first order and over a limited region, e.g. \citet[][and refereces therein]{liseau2015}. 

\begin{figure}[H]
  \resizebox{\hsize}{!}{
    \rotatebox{270}{\includegraphics{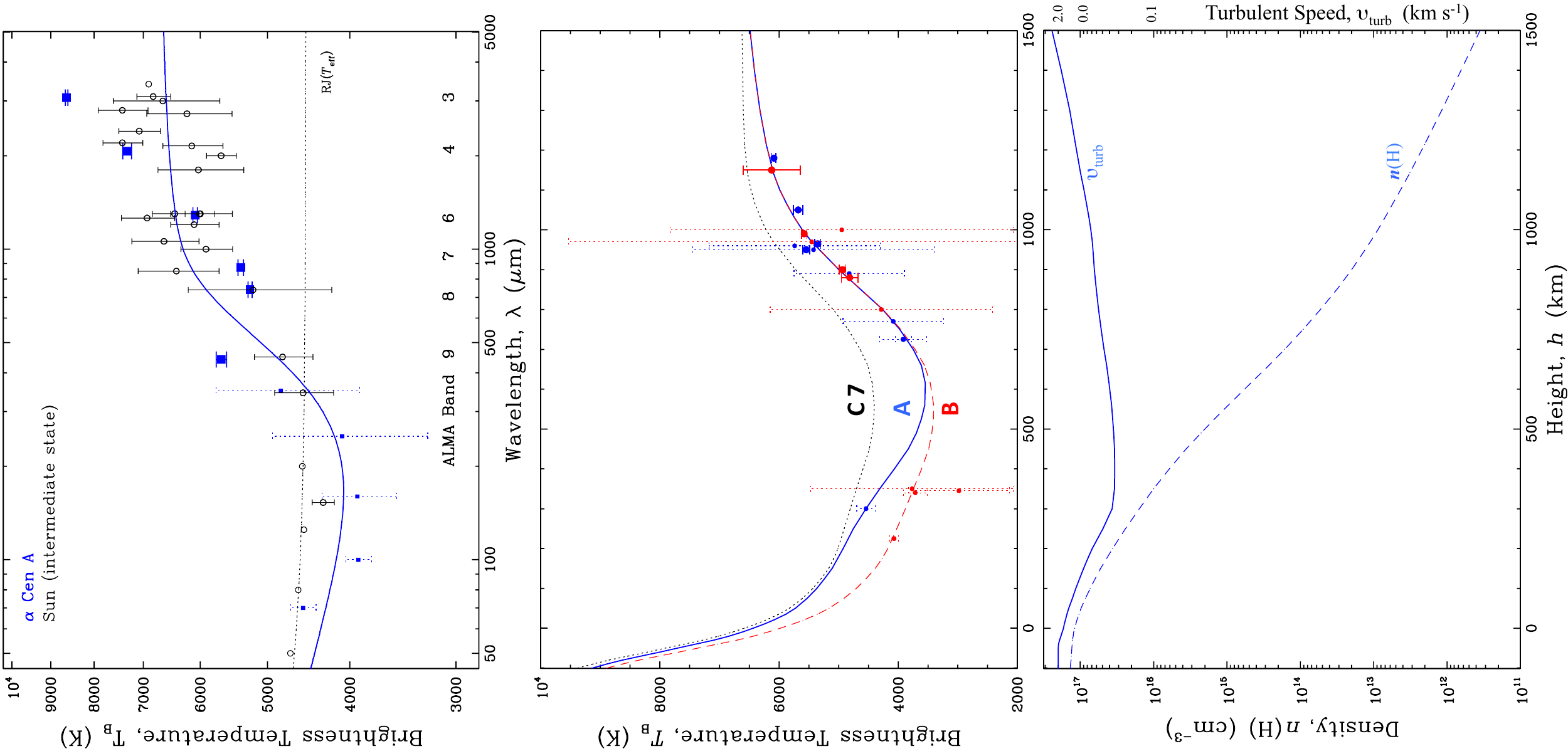}}
                        }
  \caption{{\bf Top:} The SED of the model chromosphere of the G2 star \acena, based on the modified solar C\,7 model, is shown by the blue curve.  Data are from {\it Spitzer, Herschel} and APEX \citep{liseau2013} (small blue sysmbols and dotted error bars) and from ALMA (big blue squares). The ALMA bands are indicated at the bottom of the figure and the stellar photosphere is shown as RJ($T_{\rm eff})$. For comparison, data for the Quiet Sun from \citet{loukitcheva2004} are shown as black open circles. {\bf Middle:} The run of $T_{\rm B}$ with height $h$ with symbols as above. For comparison, also the corresponding model for the K\,1 star \acenb\ is shown in red, and, for reference, the solar C\,7 model as black dots. {\bf Bottom:} The run of density $n$(H) and turbulent velocity $\upsilon_{\rm turb}$ with height $h$ is shown for the solar analogue \acena.
     }
  \label{a_sun}
\end{figure}

A more sophisticated method is to build a theoretical model chromosphere that at its base is anchored in  the photosphere. The result of this is displayed in Fig.\,\ref{a_sun}, showing both the temperature minimum and the temperature increase that are retrieved by the semi-empirical non-LTE model chromosphere of \acena, based on a modified hydrostatic equilibrium model (C7) of the solar chromosphere \citep{avrett2008,delaluz2014}. C7 can be viewed as an average of the five most widely used solar chromosphere models \citep{vernazza1981,loukitcheva2004,fontenla2007,avrett2008,delaluz2014}. 

The temperature profile is computed iteratively from the modified density/pressure structure, ionization balance and opacity (lines and continua). As the conditions in the chromosphere strongly deviate from thermodynamical equilibrium, both the ionization-excitation and the radiative transfer are treated in non-LTE (De la Luz \& Tapia, in preparation). Figure\,\ref{a_sun} also shows the sharp drop in proton density $n(\rm H)$ and the increase of the turbulent speed $\upsilon_{\rm turb}$, steepening into shocks. Although \acenb\ is not a solar analogue like A, a modified solar model also provides an acceptable fit to the data. The modeled $T_{\rm B}(h)$ of the K-star \acenb\ is also shown in Fig.\,\ref{a_sun}.   

For \acena, the temperature profile is shallower than for the Sun and $T_{\rm min}=3548$\,K at $h=615$\,km, where the proton density $n({\rm H})=4.7 \times 10^{14}$\,\cmthree. The corresponding model parameters for \acenb\  are 3407\,K, 560\,km and $9.5\times 10^{14}$\,\cmthree, respectively.  

The temperature minimum in the $T_{\rm eff}$-scale of the \acena\ model, $T_{\rm min}/T_{\rm eff}=0.61$, is as low as what has been observed in CO lines from the Sun  \citep[$T_{\rm min}/T_{\rm eff}=0.65$, ][and references therein]{avrett2003}. This is lower than what traditionally has been derived from the wings of resonance lines, viz. $> 0.7$ for both \acena\ and the Sun \citep{ayres1976,avrett2003}.

At the longest wavelengths the exponent of the observed SED changes, likely because the free-free emission is turning from optically thick to thin beyond 1.5\,mm (frequency exponent tends from about 2 to 0). Especially at 3\,mm, the Band\,3 data are not well reproduced by the model, the density of which is too low to generate sufficient  free-free and H$^{-}$ opacity for the required flux. However, from Table\,\ref{tb}, it can be seen that the radiation from \acena\ in Band\,4 and 3 probably originates rather high up, at about 2000\,km and near the base of the transition region (TR) into the hot corona, which is seen in the X-rays from the \acen\ binary \citep{dewarf2010,ayres2014}. The X-ray emission is particularly strong from the more active companion \acenb.

It is likely that it is in these thin layers of the TR base, where wave energy is dumped and dissipated \citep{soler2015,shelyag2016}. Therefore, this region is critical to the understanding of the heating processes of the outer atmospheres of the stars and the Sun. Given the available evidence, ALMA Band\,5 observations will eventually be particularly crucial for the observation of these layers in the \acen\ stars. These stars deserve continued monitoring, including observations at longer wavelengths. 
 
\acena\ and B are known to be variable on both short and long time scales \citep{dewarf2010,ayres2015}. In X-rays and the FUV, both stars show flickering but also solar-like magnetic cycles, with \acenb\ being the more active one. Repeat observations would assess the level of activity in the submm/mm regime. Between 2014 and 2017, \acena\ is expected to go through its broad shallow maximum of its $\sim19$\,year cycle, whereas B will presumably pass through a minimum of its  8\,year cycle. Thus, perhaps in contrast to the solar case, changes of the chromospheric emission from the active K-dwarf could occur over a period of a few years, although such behavior, by analogy with the Sun, would not be expected for the less active \acena.  


\begin{figure}
  \resizebox{\hsize}{!}{
    \rotatebox{00}{\includegraphics{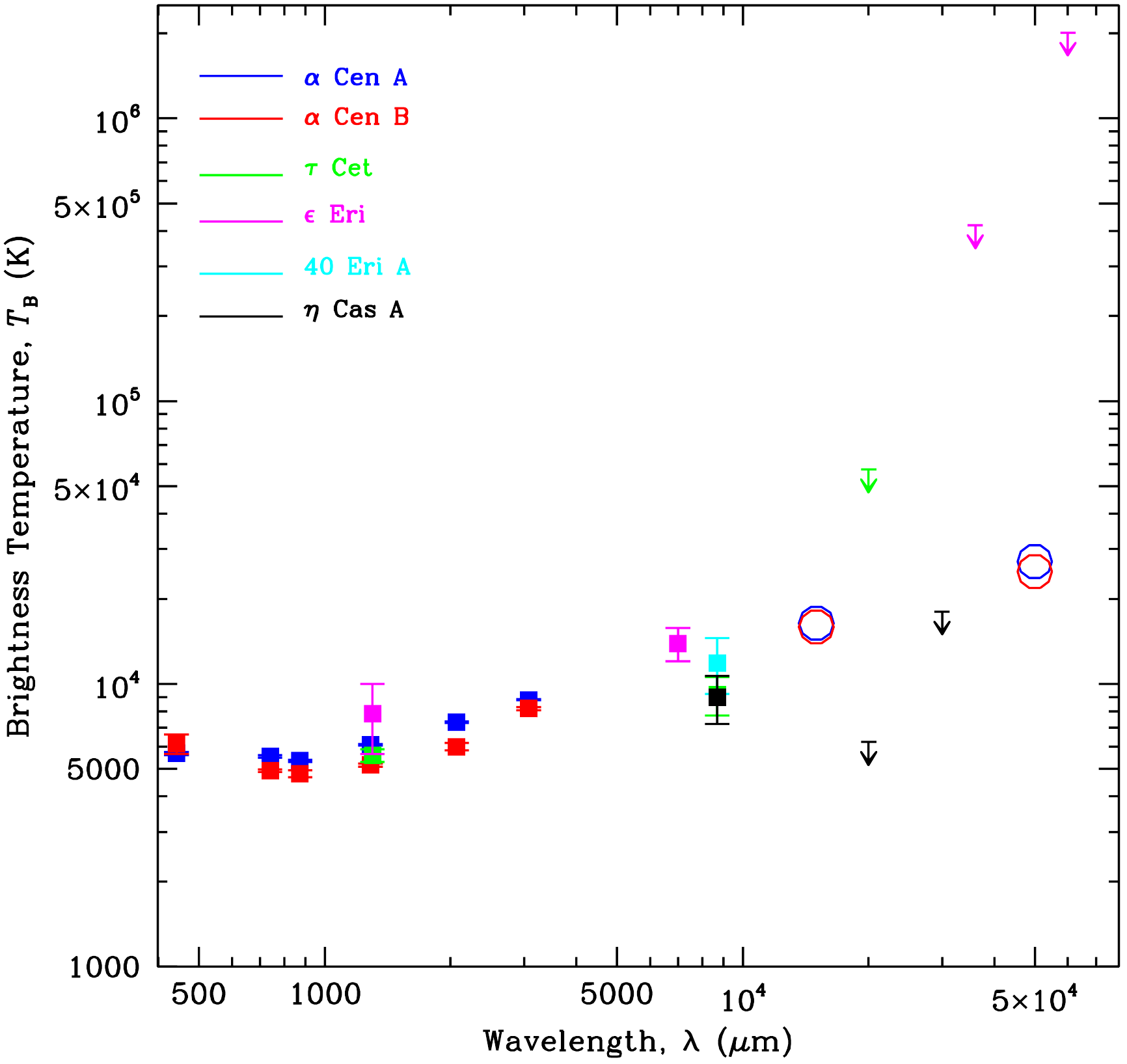}}
                        }
  \caption{Brightness temperatures for six solar-type stars at wavelengths from 0.5\,mm to 6\,cm (see the text). Detections were obtained below 1\,cm and merely upper limits above that wavelength. The color coding and stellar identifications are given in the upper left corner of the figure. The open circles denote estimates of future ATCA detections of \acena B\ in 12 hours at 20 and 6\,GHz, respectively (see the text).
   }
  \label{stars_SED}
\end{figure}

\subsection{Comparison with other stars}
\subsubsection{Solar-type}

In addition to \acen, a handful of other solar-type stars (late F to early K) have been observed at long wavelengths. These stars are all within 6\,pc. For $\epsilon$\,Eridani (K2\,V) measurements have been made at 1.3\,mm and 7\,mm \citep{macgregor2015} and at 3.6\,cm \citep{guedel1992} and 6\,cm \citep{bower2009}; for $\tau$\,Ceti (G8.5\,V) at 1.3\,mm \citep{macgregor2016} and at 8.7\,mm and 2\,cm \citep{villadsen2014}. Further, 40\,Eridani\,A (K0.5\,V) and $\eta$\,Cassiopeiae A (F9\,V) at 8.7\,mm, and the latter also at 2 and 6\,cm, have also been observed by \citet{villadsen2014}. 

As seen in Fig.\,\ref{stars_SED}, there is only limited overlap with the wavelength domain of the \acen\ binary and upper limits, rather than detections, dominate at cm-wavelengths. However, for all detected cases (4 stars in addition to \acena\ and  B), the fluxes were not consistent with photospheric values but significantly higher. Therefore, it was generally concluded that this excess emission originates in stellar chromospheres, similar to those in the Sun and \acena B.

\begin{figure*}
  \resizebox{\hsize}{!}{
    \rotatebox{00}{\includegraphics{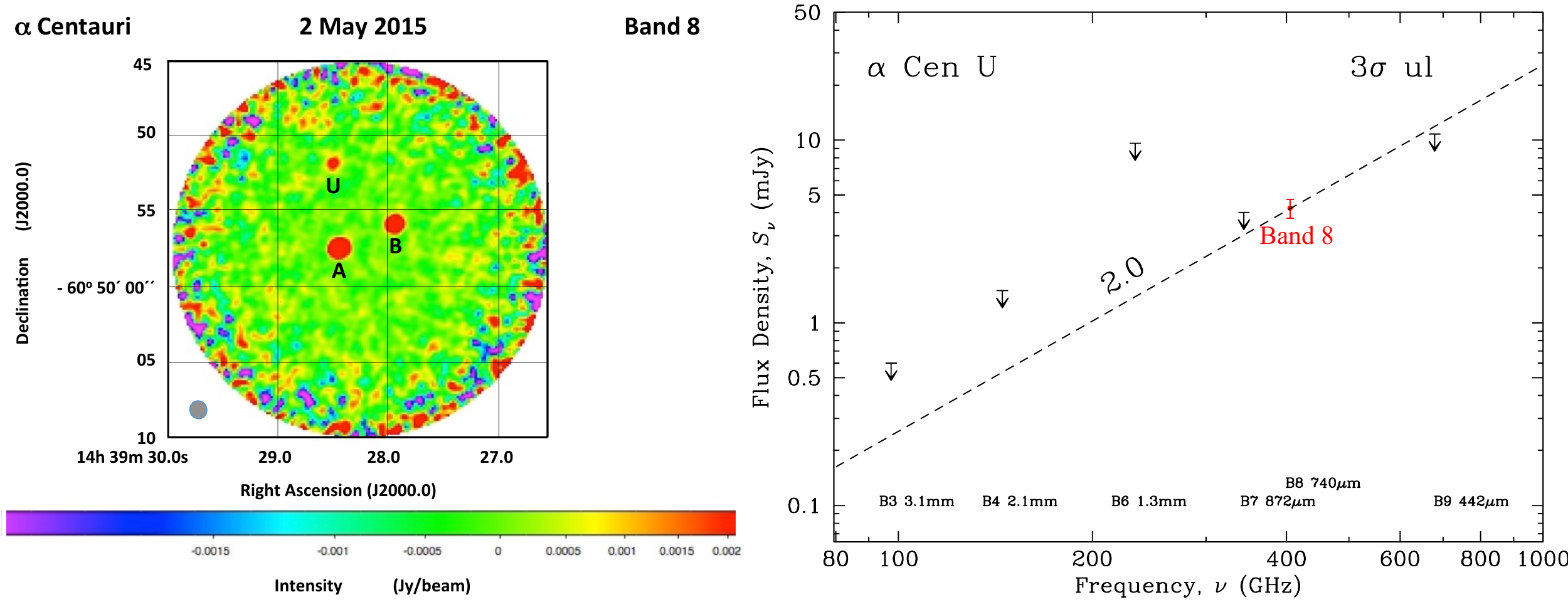}}
                        }
  \caption{{\bf Left:} Band\,8 observation of \acen tauri on 2 May 2015, with the color bar for the intensity shown below. Apart from the well known binary \acena\ and \acenb, a previously unknown source U was discovered less than 6\asec\ north of the primary A. The displayed image is primary beam corrected and the slightly oval synthesized beam (Table\,\ref{pos}) is shown in grey in the lower left corner. In the FITS image,  the FK5 (J2000.0) coordinates at mid-integration, i.e. JD\,2457144.632219328, are R.A.=\radot{14}{39}{28}{491},  Dec.=\decdot{$-60$}{49}{51}{83}.
{\bf Right:} The logarithmic submm/mm-SED of the unidentified source U near \acen\ is consistent at the $3\,\sigma$ level with that of a blackbody, as indicated by the dashed line of slope 2.0 (cf. Table\,\ref{Udata}).
   }
  \label{U_Band8}
\end{figure*}

\subsubsection{Giants}

\citet[][and references therein]{harper2013} discuss ongoing observational and theoretical work on giants (luminosity class III), addressing the possibility to observationally separate acoustic from MHD heating processes in the upper atmospheres due to the large scale heights in these stars. Their convective cells and envelopes are much larger than those of main-sequence stars, which may make possible to observationally distinguish between these effects. In addition and in contrast to the smaller and more compact main-sequence stars (class V), giants are relatively bright and hence offer themselves as possible candidates for calibration purposes for observations in the submm/mm/cm regime \citep[see also][]{cohen2005}.


\subsection{A new, unidentified point-like source near \acen}

In May 2015, an unidentified object was detected in the Band\,8 observations of \acen\ (Fig.\,\ref{U_Band8}). This point source, designated U and with integrated flux over the band of about 4\,mJy (Table\,\ref{Udata}), was within a few arcseconds of the binary. As this object was not detected in any other data set \citep[including UV, VIS and NIR with HST and VLT, see][]{kervella2006,kervella2007}, other epoch data are lacking and hence its nature is unknown. 

Figure\,\ref{U_Band8} also displays the SED of this object, consisting of one detection and five upper limits at the $3\sigma$ level. However, the data could be consistent with blackbody emission, viz. $S_{\!\nu}\propto \nu^2$, and may be due to a submm galaxy, a stellar object, a brown dwarf or a planetary object. A companion star of the \acen\ system does not present a viable explanation, as any star would be brighter than 10th magnitude in the V-band, and hence must be discarded.

The submm galaxy option would imply that the proper motion of U would be minuscule, and that it would be quickly left behind the \acen\ stars as they pace, at the rate of \asecdot{3}{7}\,yr$^{-1}$, through the sky. As \acen\ is in close projection to the plane of the Galaxy, a stellar nature of U may perhaps appear more natural. However, this putative star remained undetected in recent deep searches, implying that U is either a distant heavily extinguished background star or a nearby, very cold object, i.e. a brown dwarf or a planetary object. The parallax and proper motion would clearly distinguish among these possibilities. 

Verly low-temperature brown dwarves like the T\,8.5-type ULAS\,$J003402.77-005206.7$ with an estimated temperature of 575\,K, or the even cooler Y\,2 object  WISE\,J$085510.83-071442.5$ with  $T_{\rm eff}=250$\,K \citep{tinney2014,leggett2015}, can serve as known examples, i.e. an extremely cool brown dwarf at a distance of nearly 20\,000\,AU may be a viable candidate for the identification of source U. However, like the Y2 object, the Wide-field Infrared Survey Explorer (WISE) should have picked it up. Unless close to the very bright \acena B, the moderate angular resolution of WISE ($>$\asecdot{6}{0}) presented an obstacle to a clean detection. 

\begin{figure*}
  \resizebox{\hsize}{!}{
    \rotatebox{00}{\includegraphics{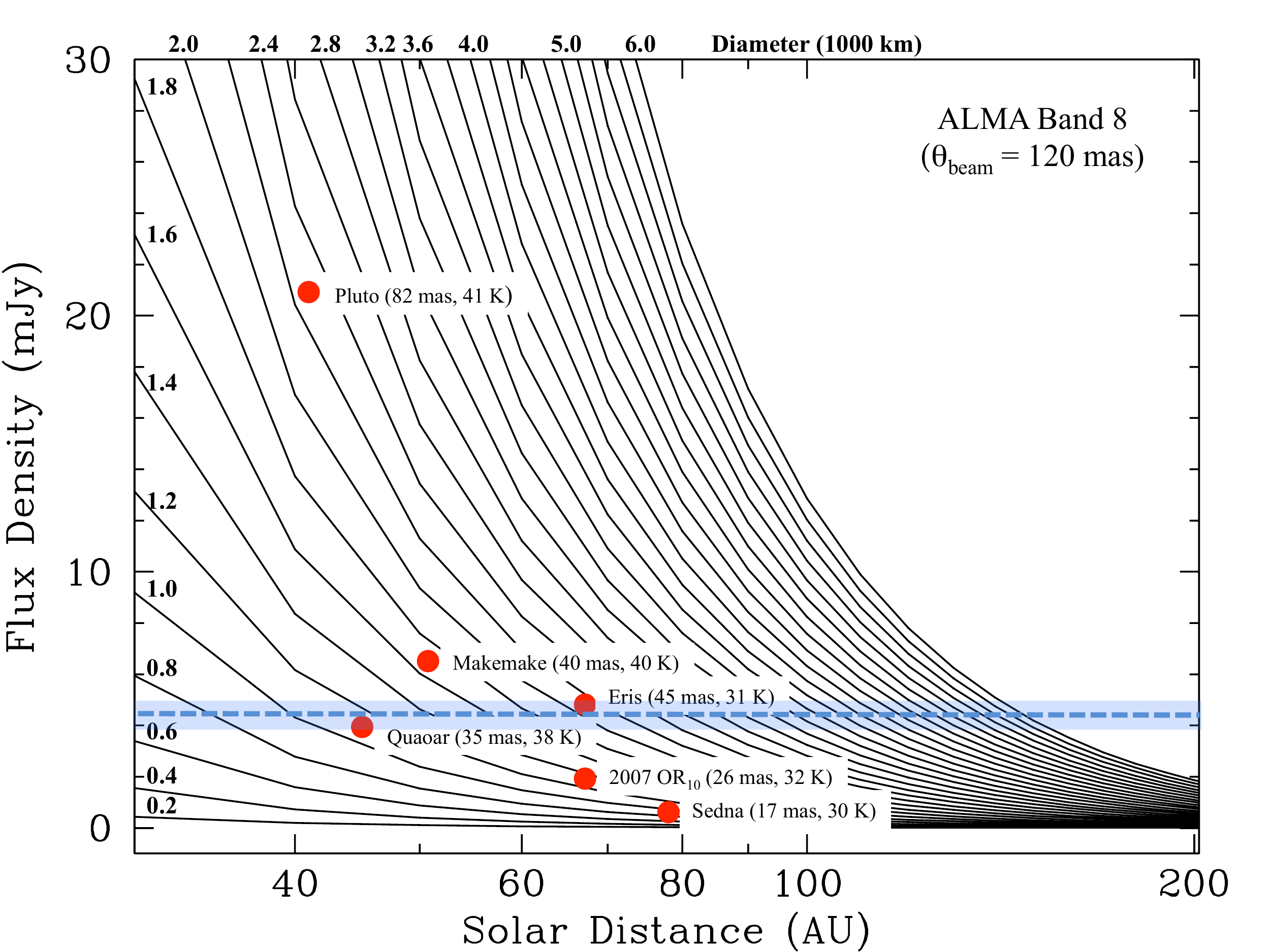}}
                        }
  \caption{Band\,8 flux density as function of the distance from the Sun with diameters as parameter, in \powten{3}\,km and next to or atop the curves and arbitrarily limited to 6000\,km, i.e. slightly smaller than the diameter of Mars. Both the surface temperature and the radius are a priori undetermined. A few known TNOs with their names are shown by the red dots (www.minorplanetcenter.org/iau/lists/Sizes.html). In parentheses, the apparent diameter in milli-arcseconds and the estimated blackbody temperature are given. The size of the ALMA synthesized beam, $\theta=120$\,mas, is given in the upper right corner, confirming that these objects would be point-like to ALMA. The observed Band\,8 flux density of the unidentified object is indicated by the horizontal blue-shaded dashed line ($\pm 1 \sigma$). The distance to the U-source remains to be determined. 
         }
  \label{TNO}
\end{figure*}

In the solar system, the projected offset of \about\,\asecdot{5}{5} would correspond to a distance between Jupiter and Saturn\footnote{The accuracy of the absolute stellar positions will be addressed by Kervella et al. 2016 (in preparation)}. However, the identification of U as a planetary companion of \acen\ would be totally unrealistic, because the observed 740\,\um-flux would be too high by several orders of magnitude. If a body of planetary dimensions, U would possibly be bound to the solar system, but its distance would presently be undetermined. Fig.\,\ref{TNO} shows the distances and flux densities at 740\,\um\ estimated for several known dwarf planets with the diameter as parameter. From the figure it is evident that U is likely more distant than Pluto, since an \about\,1000\,km body at roughly 40\,AU would have been known for a long time, i.e. for at least ten years. For example, when examining a total of 766\,925 known solar-system objects\footnote{http://www.minorplanetcenter.net/cgi-bin/mpcheck.cgi} for being within 15\amin\ around \acen\ at the time of observation, we found no source down to the limiting $V$-magnitude of 26.0. Therefore, a low-albedo, thermal Extreme Trans Neptunian Object (ETNO), would clearly be consistent with our data (see Fig.\,\ref{TNO}). 

\begin{table*}
\caption{Primary beam corrected flux density and $1\sigma$ upper limits for the U-source in mJy}  
\label{Udata}      
\centering          
\begin{tabular}{c ccccc }     
\hline\hline \\ 
\smallskip
\smallskip                                                                                  		                    
 Band 9   	& Band 8               	& Band 7  			& Band 6    		&  Band 4    	& Band 3           \\         
679\,GHz	& 405\,GHz		& 343.5\,GHz		& 233\,GHz 		& 145\,GHz	& 97.5\,GHz	 \\
442\,\um 	& 740\,\um		& 873\,\um 		& 1287\,\um 		& 2068\,\um	& 3075\,\um	\\
\hline \\
$<3.6$	&  $4.24\pm 0.49$    & $<1.34$			& $< 3.2$			& $< 0.5$		 & $< 0.2$ 	\\
\hline                
\end{tabular}
\end{table*}

\section{Conclusions}
 
 Below, we briefly summarize our main conclusions.
 
 \begin{itemize}
\item[$\bullet$] ALMA observations of \acen tauri at 0.44, 0.74, 0.87, 1.3, 2.1 and 3.1\,mm clearly resolved the binary, but not the stellar disks, at all wavelengths. The spectral energy distributions of these continuum measurements are consistent with radiation that follows $S_{\!\nu} \propto \nu^2$, except at the lowest frequencies where the SEDs appear to flatten. This is particularly pronounced for the more active secondary, a K\,1 star, possibly indicative of time variability within half a year or, perhaps more likely, of optically thin free-free emission.   
\item[$\bullet$] The ALMA data have been modeled with modified solar chromosphere models which result in the physical structure of the stellar chromospheres. This adapted solar model works very well for the solar analog \acena\ (G2\,V), but also for the K1\,V star \acenb. Comparison with the data indicates that the temperature minima of both \acena\ and B are deeper than on the Quiet Sun. These correspond to the low temperatures seen in lines of the CO molecule on the Sun and occur at atmospheric heights of 615\,km and 560\,km, respectively.  
\item[$\bullet$] The ALMA data for \acena B can be put into context with observations of other nearby solar-type stars that show that chromospheric mm-wave emission is a common feature among these stars and that an increase in the sample size can be expected in the near future.
\item[$\bullet$] The ALMA imaging at 0.74\,mm led to the discovery of a previously unknown point source within a projected distance of 7.5\,AU from \acena B. The ALMA observations were performed at different occasions during one year (2014 - 2015), but this source was clearly detected only on one date. At the three sigma level, the SED of this object is consistent with that of a blackbody and we speculate about its nature. Unless it is a highly variable background source, we find it most likely that it is a distant member of our solar system.
\end{itemize}

\begin{acknowledgements}
We thank the referee for his/her valuable comments on the manuscript. This paper makes use of the following ALMA data: ADS/JAO.ALMA\#2013.1.00170.S. ALMA is a partnership of ESO (representing its member states), NSF (USA), and NINS (Japan), together with NRC (Canada) and NSC and ASIAA (Taiwan), in cooperation with the Republic of Chile. The Joint ALMA Observatory is operated by ESO, AUI/NRAO, and NAOJ.
\end{acknowledgements}

\bibliographystyle{aa}
\bibliography{aCen_allALMA}

\begin{appendix}
\section{Sub-band fluxes}

\begin{figure}
  \resizebox{\hsize}{!}{
    \rotatebox{00}{\includegraphics{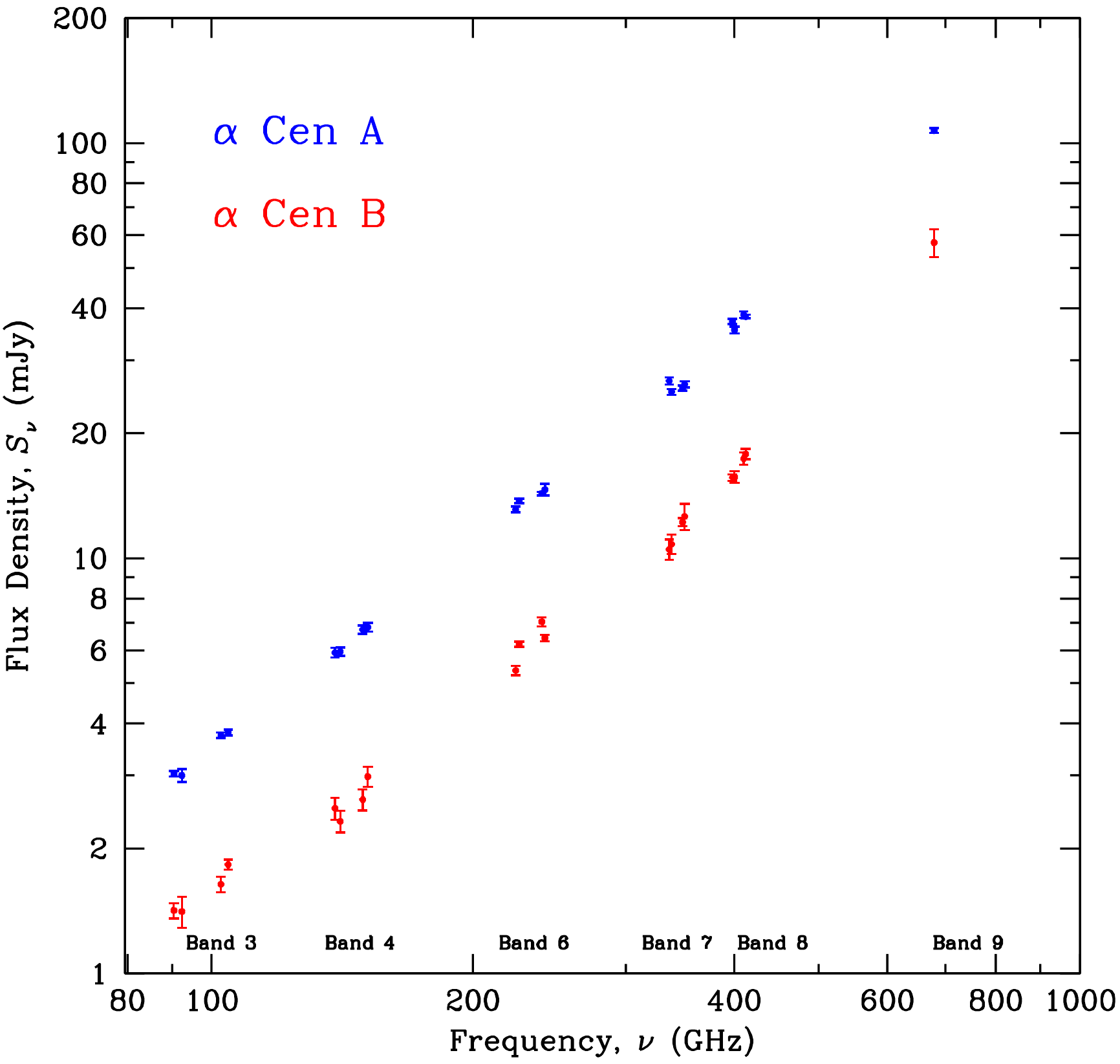}}
                        }
  \caption{Measurements of the flux density of \acena\ (blue circles) and of \acenb\ (red circles) in the sub-band windows (spw), see Table\,\ref{A_sw}. The error bars represent the $1\sigma$ rms-values. Band\,9 is too narrow to allow meaningful measurement in sub-windows and only a single value is given.}
  \label{flux_sw}
\end{figure}

The flux densities of the spectral windows per band are provided in Table\,\ref{A_sw} (\acena) and the data are plotted in Fig.\,\ref{flux_sw}. For Band\,9, only a single value is given, as the windows are too narrow for meaningful individual measurement.

For \acenb, the relative drop in intensity in the second spw of Band\,4 is conspicuous. This is not evident for \acena, and the glitch can therefore not be caused by different calibrations. \acena B are point sources and were observed simultaneously. Hence, simultaneous visibility fitting, with fixing the positions to reduce the noise \citep{marti-vidal2014}, should not result in such large differences, unless there is something in the data, e.g. a spectral feature, in \acenb\ that is not present in the SED of \acena. New observations of B, at higher S/N in Band\,4, would be necessary to resolve this issue.

\begin{table}
\caption{Sub-band (spw) flux densities for \acena B}             
\label{A_sw}      
\centering          
\begin{tabular}{ccr c rl }     
\hline\hline \\ 
\smallskip
\smallskip     
B     &	$\nu$    &  $S_{\nu}$(A)    & rms(A)   &$S_{\nu}$(B) & rms(B)	\\
         &  (GHz)	    &        (mJy)     	& (mJy)	&  (mJy) 		&   (mJy) \\
\hline    \\           
3  &	\phantom{1}90.49459 &	3.03 	&	0.04 & 1.42 		& 0.06\\	
3  &	\phantom{1}92.43209 &	3.00 	&	0.11 & 1.41 		& 0.12\\
3  &	102.4946 			  &	3.75 &	0.06 	& 1.64	 	& 0.07\\
3  &	104.4946 			  &	3.81 &	0.06 	& 1.83 		& 0.05\\
\\
4  &	138.7133 			&	5.92	& 	0.15	 & 2.50		& 0.15\\
4  &	140.6508 			&	5.96 	&	 0.14	 & 2.32		& 0.14\\
4  &	149.2758 			&	6.74 &	 0.15	 & 2.62		& 0.15\\
4  &	151.2758  		&	6.82 &	 0.16	 & 2.98		& 0.16\\
\\
6  &	224.000   			&	13.12 &	0.22 	& 5.37		& 0.14 \\
6  &	226.000   			&	13.75 &	0.17 	& 6.21		& 0.09\\
6  &	240.000   			&	14.33 &	0.14 	& 7.03		& 0.18\\
6  &	242.000   			&	14.64 &	0.46 	& 6.43		& 0.11\\
\\
7  &	336.4946 			&	26.75 &	0.53 	& 10.50		& 0.59\\
7  &	338.4321 			&	25.18 &	0.39 	& 10.81		& 0.58 \\
7  &	348.4946 			&	25.69 &	0.38 	& 12.22		& 0.27\\
7  &	350.4946 			&	26.25 &	0.48 	& 12.61		& 0.90 \\
\\
8  &	397.9946			 &	37.17 &	0.57	 & 15.66		& 0.28\\
8  &	399.9321 			&	35.47 &	0.65	 & 15.69		& 0.51\\
8  &	409.9946 			&	38.65 &	0.65 	 & 17.39		& 0.57\\
8  &	411.9946 			&	38.19 &	0.38 	 & 17.84		& 0.51\\
\\
9  &	678.9600  		&	107.20 &	1.50 	 & 57.60		&  4.50\\
\hline  
\end{tabular}
\end{table}

\end{appendix}

\end{document}